\documentclass[lettersize,journal]{IEEEtran}
\usepackage{amsmath,amsfonts}
\usepackage{algorithmic}
\usepackage{array}
\usepackage[caption=false,font=normalsize,labelfont=sf,textfont=sf]{subfig}
\usepackage{textcomp}
\usepackage{stfloats}
\usepackage{url}
\usepackage{verbatim}
\usepackage{graphicx}
\def\BibTeX{{\rm B\kern-.05em{\sc i\kern-.025em b}\kern-.08em
    T\kern-.1667em\lower.7ex\hbox{E}\kern-.125emX}}
\usepackage{balance}
\begin{document}
\title{Manufacturing Tolerances of Non-Planar Coils for an Optimized Tabletop Stellarator}
\author{Pedro F. Gil, Vitali Brack, Tristan Schuler, Paul Huslage, E. V. Stenson}

\markboth{IEEE Transactions on Applied Superconductivity}{}%

\maketitle

\begin{abstract}
Stellarator coils are known for their complexity and departure from planarity, along with tight manufacturing tolerances in order to achieve the target magnetic field accuracy. 
These requirements can lead to increased costs and delays in assembly; failure to meet them can compromise the stellarator's performance. 
Small-scale experiments offer opportunities to develop and benchmark stellarator coil design and evaluation methods more quickly and at lower budget.
In this work, we analyze precise 3D scans of the manufacturing deviations of two 3D-printed coil frames (steel, Ti alloy) and one CNC-machined coil frame (Al alloy), as part of assessing these approaches to fabricating high-temperature superconducting (HTS) coils for a tabletop stellarator. 
The deviations are measured along the coil length, then modeled using Gaussian processes to extract characteristic length scales. 
Finally a statistical study of field accuracy is performed using relevant experimental parameters.
We conclude that the manufacturing perturbations along the winding path from CNC-machining are almost an order of magnitude lower than those from Additive Manufacturing. Together with high overall fabrication accuracy, this allows for higher magnetic field precision and an improved assembly process.
\end{abstract}


\section{Introduction}

Stellarators are magnetic confinement devices typically designed to contain high-temperature plasmas in a toroidal shape with the goal of achieving controlled nuclear fusion. 
Unlike its main alternative, the tokamak, a stellarator generates the three-dimensional magnetic field required for plasma stability and confinement almost entirely with external electromagnetic coils. 
Multiple companies consider this to be a commercially viable concept for fusion power plants \cite{typeone, proxima}.

Modern stellarator coils are usually optimized to have non-planar geometries, however, and this presents significant engineering challenges; the manufacturing of these coils and their precise positioning are demanding tasks. 
Due to high degrees of complexity together with tight tolerances, the National Compact Stellarator Experiment (NCSX) was cancelled in 2008 after budget and timeline overruns \cite{ncsx}. 
The manufacturing and assembly of the Wendelstein 7-X (W7-X) coils were similarly challenging, with tolerances of 2 mm (local deviation) in a device with 5-m major radius \cite{KLINGER}. 
Even small departures from the ideal, as-designed coil shapes and positions can introduce error fields—unwanted perturbations in the magnetic field structure. 
These error fields can degrade plasma confinement, drive instabilities, and ultimately prevent the device from achieving its performance targets \cite{pedersen}.

\begin{figure}
    \centering
    \includegraphics[width=0.9\linewidth]{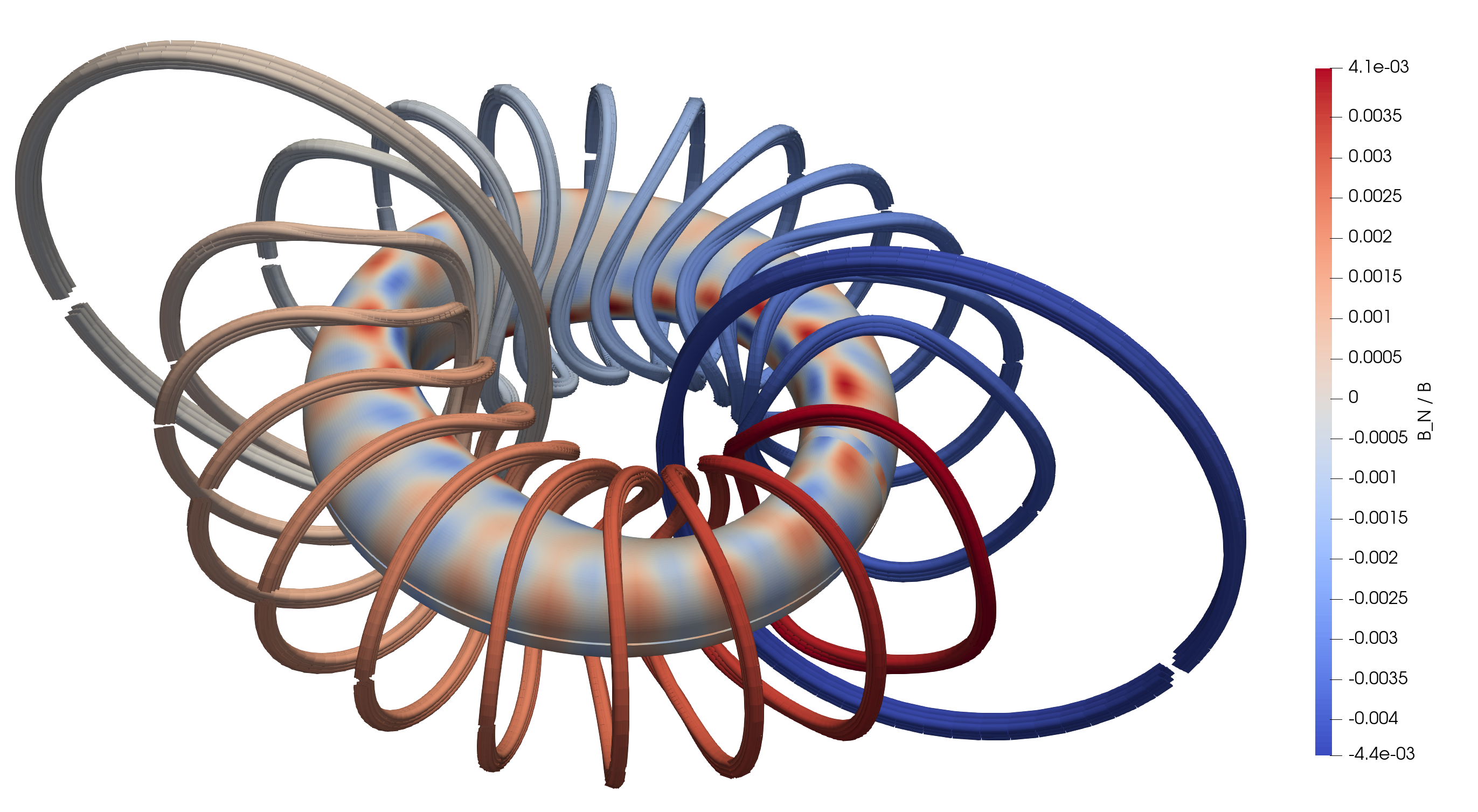}
    \caption{Coil shapes and outer plasma surface for the EPOS stellarator, as designed. The magnetic field error, due to coil ripple, is low, which would enable more than satisfactory confinement of the low-energy electrons and positrons for which it is intended.}
    \label{fig:epos}
\end{figure}

While fusion stellarators require coils that are several meters in scale, small-scale experiments to investigate fundamental plasma science can offer (alongside their physics goals) ``rapid prototyping'' opportunities for developing and benchmarking improvements in design and characterization of coil shapes.
The new, tabletop-sized stellarator EPOS (Electrons and Positrons in an Optimized Stellarator) is currently going through design reviews, in preparation for being built at the Max Planck Institute for Plasma Physics (IPP) in Garching, Germany; its purpose is to confine a matter-antimatter ``pair plasma'' \cite{epos}. 
The plasma will have a major radius of around 20 cm and minor radius of about 4 cm (Figure \ref{fig:epos}).
The coils will have diameters in the 20-50 cm range and will be made of 3-mm-wide high-temperature superconducting (HTS) tape wound around metal frames. 
They will be conductively cooled in UHV to cryogenic temperatures \cite{paul_non_planar}.

EPOS's small dimensions imply that the magnetic field could be very sensitive to imperfections in the manufacturing of the coils, which could in turn damage the quality of magnetic field's ``quasisymmetry'', associated with charged particle confinement \cite{boozer, nuhrenberg, helander2014, rodriguez}. 
Therefore it is essential to accurately quantify the deviations between the ideal design and the as-built reality of the coils, in the same spirit as studies done for W7-X \cite{tamara1, tamara2}. 
By doing this first with test coils, we can assess which design, manufacturing, and assembly techniques can be expected to yield the most accurate results; we can also develop strategies to either correct for or mitigate the impact of inaccuracies that do arise.
Ultimately, we will  validate the quality of the final construction. 

In this paper, we describe our methods for measuring manufacturing deviations in test coil frames produced from different fabrication methods and metal alloys (Section \ref{sec:methods}), then the modeling framework used to capture the main scale-lengths associated with uncertainties between ideal and as-built coils (Section \ref{sec:models}). 
In Section \ref{sec:results} we present the main results of the metrology, complemented by Monte-Carlo simulations of the associated error fields, then conclude with a quantification of the field degradation from the manufacturing errors.

\section{Manufacturing and measurement methods}
\label{sec:methods}

\subsection{Coil frame fabrication and preparation}
\begin{figure}
    \centering
    \includegraphics[width=1.0\linewidth]{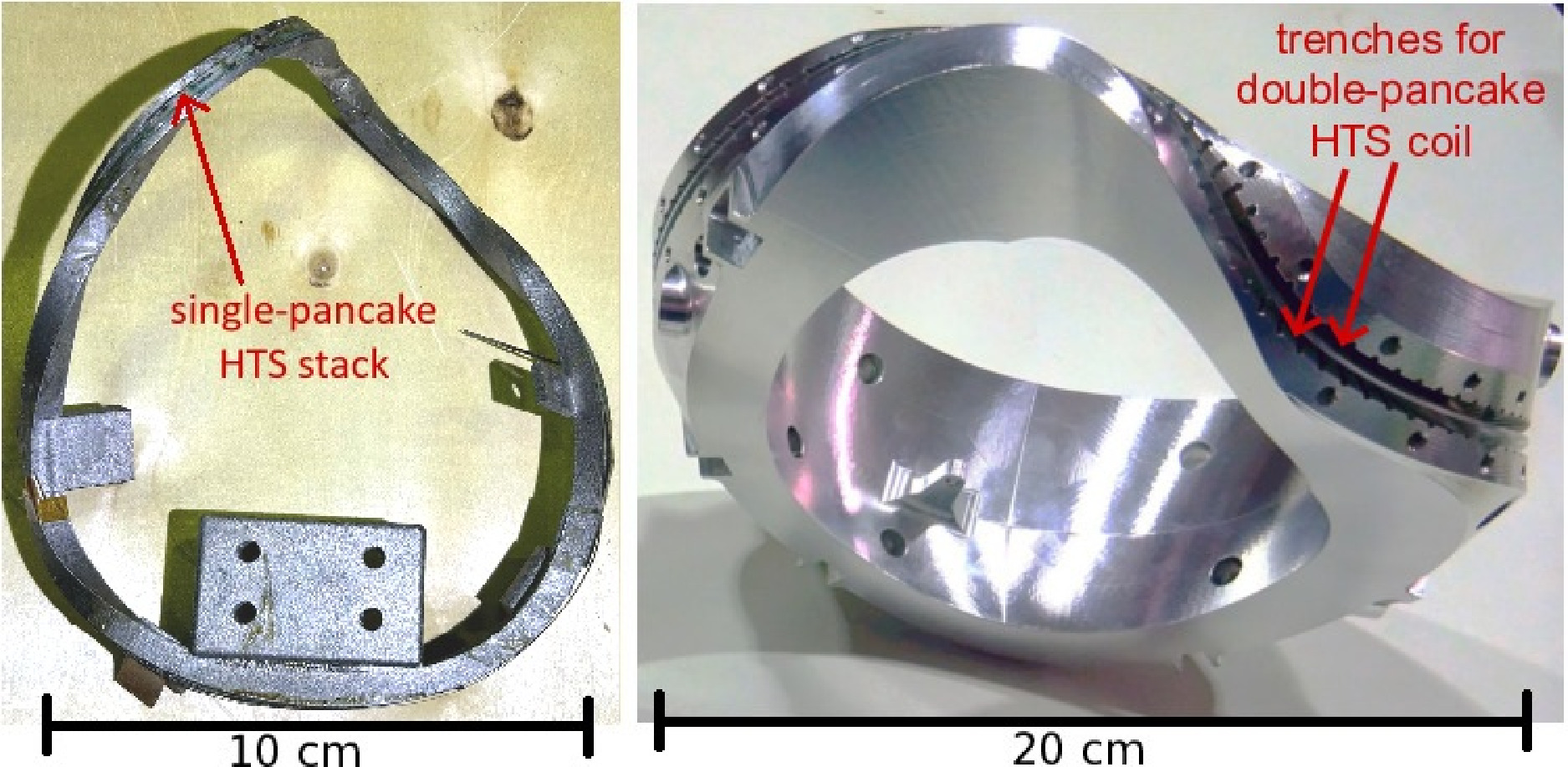}
    \caption{\textbf{Left}: 3D-printed steel coil frame, after being wound with HTS. This was an early test coil with a minimalist winding trench for a single HTS stack. \textbf{Right}: CNC-milled aluminum coil frame, with two trenches into which HTS tape would be wound. This frame is much wider, because it is designed to abut and align with its neighbors.}
    \label{fig:framephotos}
\end{figure}

In this work, we analyzed three different coil frames used to test aspects of the EPOS design. 
The coil shapes and winding packs were calculated using SIMSOPT, a stellarator optimization framework \cite{simsopt}, then exported to CAD programs for designing the winding frames around the tape stacks. 
Two identical frames were made out of steel (1.4404) and titanium (TiAl4V4) by 3D Activation using Selective Laser Melting (SLM) technology (a form of additive manufacturing (AM)). 
A third coil frame was manufactured from aluminium (7075) using a 5-axis CNC mill (DMU 100 monoBLOCK by DMG Mori) by the ITZ (Integrierte Technikzentrum) at IPP.
The steel and aluminum frames are pictured in Fig. \ref{fig:framephotos}.

Prior to scanning, the frames were cleaned with isopropanol to remove any dust particles or grease from the surfaces, then securely mounted on a designated support structure to ensure stability and prevent any movement during the measurement process. A stable temperature of about 22$^{\circ}$ C was maintained, minimizing thermal expansion or contraction.

\subsection{Data acquisition and processing}
The geometric data was acquired using the Hexagon Absolute Scanner AS1. 
This system operates by projecting a laser line onto the target surface, capturing its reflection with a high-speed camera, and employing standard laser triangulation \cite{lasertriang, lasertriang2}. 
The scanner has an accuracy of a few tens of micrometers, making it suitable for high-precision tolerance verification. 
It was mounted on a 7-axis Absolute Arm, which allowed for precise positioning, enabling the operator to reach nearly all surfaces of the coil's 3D geometry, including undercuts and curves. 
(The bottoms of the trenches not be consistently measured because of shadowing effects from the trench walls.)

The scanning process was done manually; the operator systematically moved the scanner head over the entire surface of the coil, maintaining an optimal distance and angle to ensure the best possible data capture. 
The acquisition software (Polyworks metrology software) provided real-time feedback, so as to verify complete coverage and ensure adequate point density. 
Multiple passes from different viewpoints were conducted to capture the full 3D geometry and minimize data shadows.

Upon completion of the scan, the raw data, consisting of millions of individual 3D coordinate points (a "point cloud"), underwent a series of processing steps.
The first was noise reduction, where a filtering algorithm was applied to remove any stray data points or statistical noise inherent in the measurement process, so as to produce an accurate representation of the coil's surface. 
Second, the remaining point cloud was aligned with the nominal CAD model of the coil using a best-fit alignment algorithm. 
Next, the software calculated the normal distance from each point in the scanned cloud to the nearest surface on the CAD model. 
This generated a 3D deviation map, which is visualized as a color plot overlaid on the coil's geometry. 
The color scale provides a detailed representation of the magnitude and location of manufacturing inaccuracies, with green typically indicating areas calculated to be closest to true, while warm (red, orange) and cool (blue/purple) colors represent positive and negative deviations, respectively. 

For the stellarator's magnetic field, it is not the overall frame precision that matters but the accuracy of the 3D coil path. 
We therefore also took a series of targeted measurements on both sides of the trenches.


\begin{figure}
    \centering
    \includegraphics[width=0.9\linewidth]{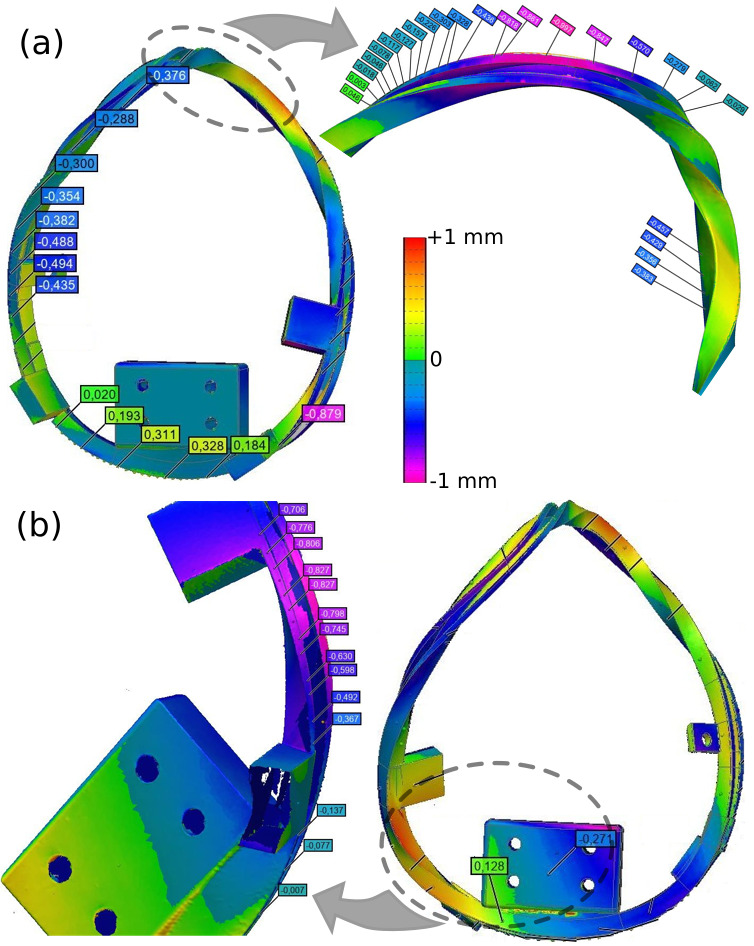}
    \caption{Results of 3D scans of the 3D-printed winding frames made of (a) steel and (b) titanium. The color scales overlaid on the CAD model indicate where and by how much (i.e., up to 1 mm) the surfaces calculated from the scans deviated from the model. In the zoomed-in segments, a series of measurement points along the sides of the coil trenches are marked.}
    \label{fig:3Dprintedframescans}
\end{figure}

\begin{figure}
    \centering
    \includegraphics[width=1.0\linewidth]{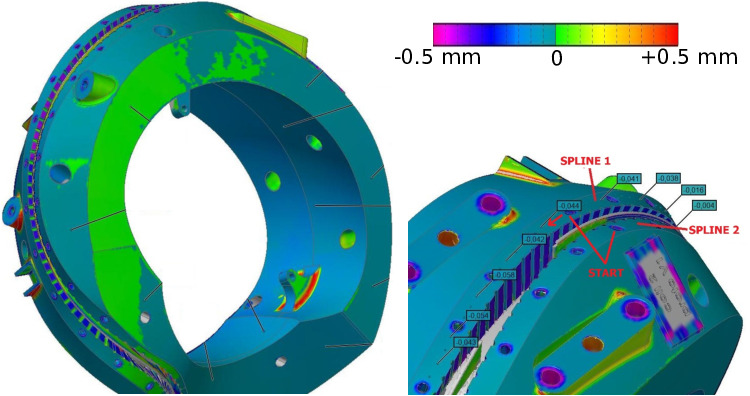}
    \caption{Results of 3D scans for the CNC-milled winding frames made of aluminum, which has smaller deviations (i.e., up to 0.5 mm), as compared to the other two frames.}
    \label{fig:CNCframescans}
\end{figure}

\section{Modeling Methods}\label{sec:models}

\subsection{Gaussian processes}
Informed by the measurements, we would like to determine characteristic length-scales for the manufacturing deviations which can then be fed back into stellarator coil modeling and optimization \cite{wechsung}. 
The method of Gaussian grocesses (GP) has been used in the past to good effect, yielding improved optimization results.

We start by modeling the coil as a one-dimensional, current-carrying filament in 3D  with path $\gamma(l) = (X(l), Y(l), Z(l))$, with $l \in [0,1]$ the normalized length along the coil and X, Y, Z the cartesian coordinates. 
To each component of the original path, a perturbation $\epsilon_i(l)$,  represented as a GP, can be added to account for manufacturing errors:
\begin{equation}
    \Tilde{\gamma_i}(l) = \gamma_i(l) + \epsilon_i(l),
\end{equation}

A GP is a generalization of random variables to functions, here the function is represented as one of the cartesian coordinates $\gamma_i(l)$. The GP is described by a kernel $k(d)$, where $d=l_i-l_j$, and $l_i$ and $l_j$ are points along the coil. From this it is possible to create a correlation function determining the smoothness and profiles of the perturbations along the coil. In this work, the following RBF kernel responsible for the correlation matrix with an extra noise term to account for experimental data variation is used:

\begin{equation}
    k(d) = \sigma^2 \exp{\left(\frac{-d^2}{2L^2}\right)} + \mathcal{N}
\end{equation}

here $\sigma>0$ captures the amplitude of the perturbations and $L>0$ determines the characteristic length over which they occur, $\mathcal{N}$ represents the noise of the signal as an independent and identically normally-distributed variable. 

\subsection{Magnetic field error estimation}
Given a model of perturbations from the ideal coil design, we can calculate and analyze what corresponding magnetic field errors would then be generated. 
In order to perform a statistical study of the field error, N samples corresponding to N different perturbed stellarators are drawn.
The kernel found by analyzing the deviations using the GP method serves as a generation tool  

The field accuracy is calculated using the quadratic flux metric. 
It is a quantity for the precision at which a coil set reproduces the intended stellarator magnetic field at the outmost boundary of the plasma:

\begin{equation}
    f_{SF} = \frac{1}{2}\frac{\int_{S} \vert \textbf{B} \cdot \textbf{n}\vert^2dS}{{\int_{S}\vert \textbf{B} \vert ^2}dS},
    \label{eq:squared_flux}
\end{equation}
where S here is the surface of the ideal target plasma, $\textbf{B}$ is the magnetic field produced by the coils, calculated with the Biot-Savart law, and $\textbf{n}$ is the normal to the surface.

\section{Results and Discussion}
\label{sec:results}

A main aim of this study is to assess different  design and manufacturing options, then use these to inform engineering choices, with the goal of building a sufficiently precise stellarator. 
The accuracy of the trenches where the HTS tape is stacked is the most important criterion to reproduce the target magnetic field; additionally, the overall structure of the frames will need to be evaluated for stellarator assembly purposes. 
Moreover, detailed scans can enable identification of systematic manufacturing errors. 




\begin{figure}[ht!]
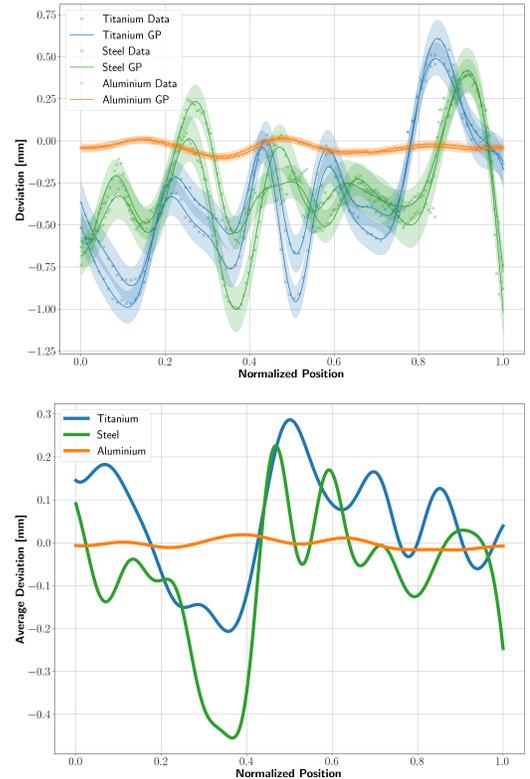

    \centering
    \includegraphics[width=0.8\linewidth]{deviation_measurements_stellarator_coils_full.png}
    
    \includegraphics[width=0.8\linewidth]{different_line1_line2.png}
    \caption{\textbf{Top}: Manufacturing deviations as a function of the normalized arc-length along winding-trench-adjacent splines for the three different coils frames. \textbf{Bottom}: Average deviation between the two splines (one on each side of each trench).}
    \label{fig:gp_deviations}
\end{figure}

Figures \ref{fig:3Dprintedframescans} and \ref{fig:CNCframescans} show the results from the 3D scans for the titanium, steel and aluminium test coil frames. 
The maximum deviations from the ideal object are about 50\% smaller in the wide, machined coil frame than the two narrow, 3D-printed coil frames. 
Note in Figure \ref{fig:CNCframescans} that along the trench walls of the aluminium coil frame there is a constant deviation of -0.3 mm; this came from a milling head misadjustment, due to human error. 
Otherwise, overall the surface finishing of the milled frame has a precision of $\pm$ 10 $\mu $m. 
For both 3D-printed coils, deviations reach up to 1 mm. 
Examples can be found in different sections of the frame, including close to the winding path and on extra structures such as the support squares for the current feeds. 
This is observed for both titanium and steel, and is above specified and standard tolerances of metal 3D printing \cite{3dmetalprinting} -- perhaps due to the high degree of complexity of this frame design or of stellarator coils in general. 

To quantify winding path inaccuracies, data points were collected from both sides of each coil frame's trench (i.e., pairs of splines), as illustrated in part by the series of points marked in the zoomed-in sections of the figures).
The deviations between the model and scan at along these splines are plotted in Figure \ref{fig:gp_deviations}. 
The start positions for the coil paths were arbitrarily chosen for the two designs; it is the same for the two 3D-printed coils. 
We see both of these frames exhibited similar errors at around the coordinate 0.9, possibly due to systematic perturbations from the manufacturing process itself. 
(For 3D printing, such errors can arise due to gravity depending on the orientation at which the objects are placed.) 
From these plots it is clear that the deviations from CNC-milling are about 10 times smaller than from 3D-printing. Moreover, if we compare the precision of both splines at the same coordinate within each coil, maximum discrepancies of 0.2 mm and 0.3 mm for the steel and titanium coils respectively are visible. For the CNC-milled part, this difference is less than 0.05 mm. 

\begin{table}
\begin{tabular}{lccc}
\hline
                        & 3D-Pr. Ti & 3D-Pr. Steel & CNC Al \\ \hline
Average GP $\sigma$ {[}mm{]}   & 2.4                 & 1.5                & 0.1                  \\ \hline
Average GP L {[}mm{]} & 42                  & 42               & 64                   \\ \hline
Average Noise Level {[}mm{]}    & 4.8e-2              & 5.6e-2           & 1e-2                 \\ \hline
Max Dev. [mm] & 0.99 & 1.02 & 0.1\\ \hline
Max Relative Dev. [mm] & 0.28 & 0.46 & 0.02\\ \hline
\end{tabular}
\caption{GP parameters for the different coils. Relative Dev. refers to deviation between the two splines on a single coil.}
\label{tab:gp_table}
\end{table}

\begin{figure}
    \centering
    \includegraphics[width=0.53\linewidth]{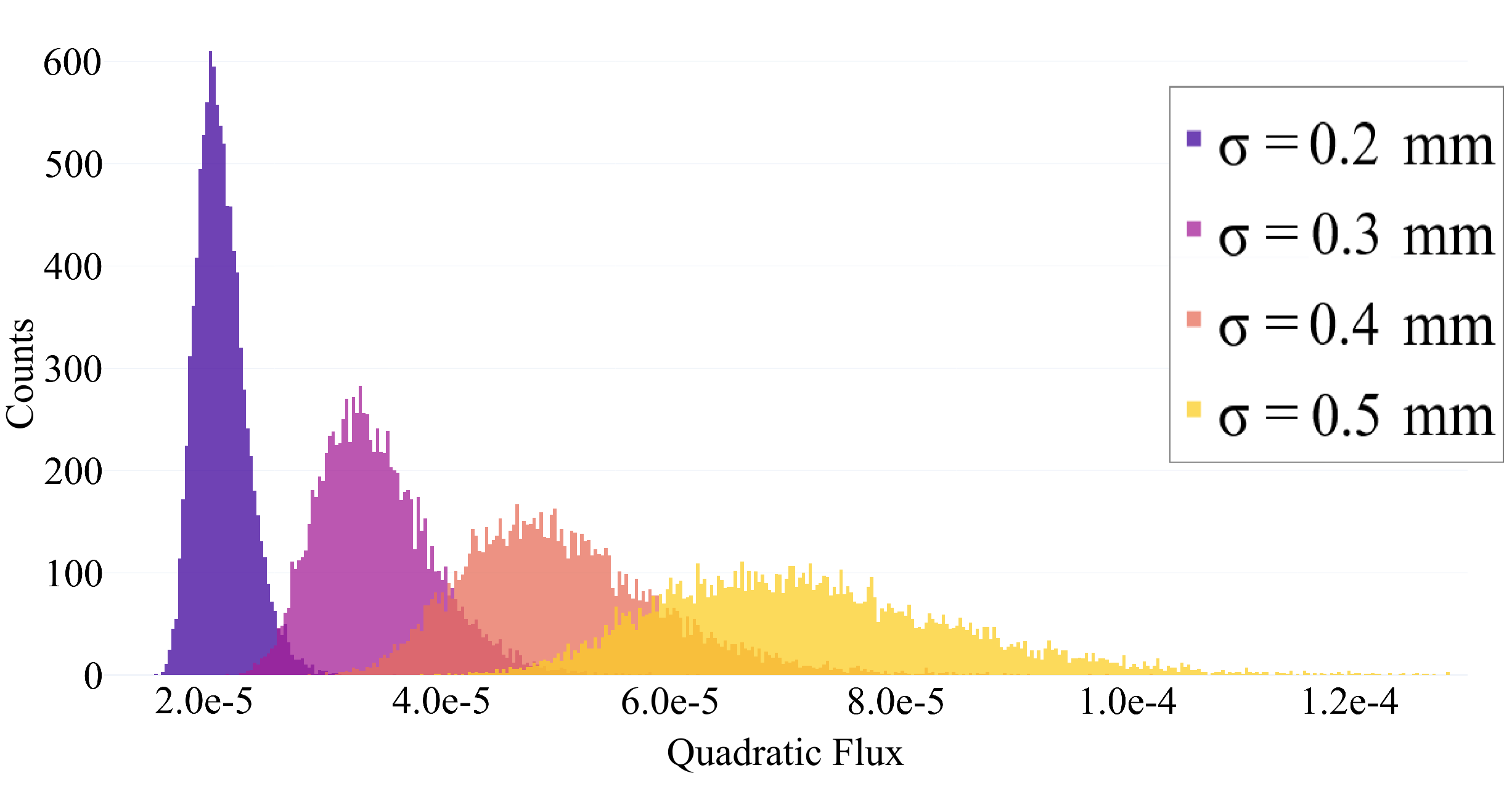}
    \includegraphics[width=0.45\linewidth]{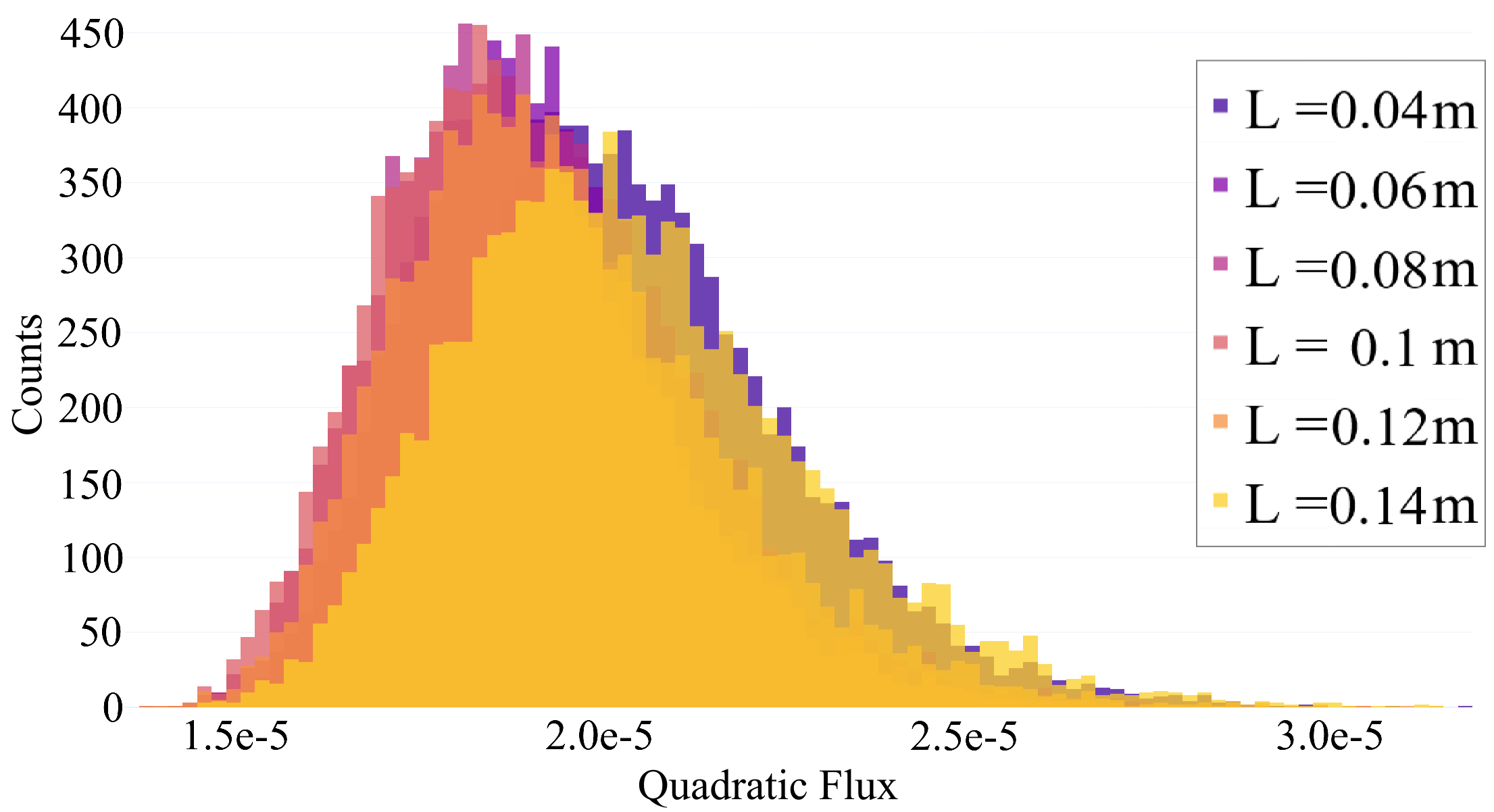}
    \caption{\textbf{Left}: Field accuracy distributions for 10 000 stellarator coil sets, perturbed with four different GP amplitudes. \textbf{Right}: Similar plot, for different characteristic perturbation lengths; unlike the amplitude distributions, changing the length does not appear to impact field accuracy.}
    \label{fig:distributions}
\end{figure}

Table \ref{tab:gp_table} summarizes the parameters obtained from fitting GPs to the splines. 
In all the parameters, the CNC-milled frame outperforms the 3D-printed frames, including noise levels (meaning that surface roughness is minimal). 

To quantify the effect of such deviations on a stellarator's magnetic field, a coil set was perturbed 10 000 times using a similar GP, mimicking various types of manufacturing defects. 
Figure \ref{fig:distributions} plots histograms of the results; each ``count'' refers to one perturbed stellarator. 
GP parameters $\sigma$ and $L$ were varied within the measured range to visualize how such parameters can impact the field accuracy (as defined in Equation \ref{eq:squared_flux}). 
Varying the perturbation length scale $L$ does not appear to modify the distribution. 
However, increasing $\sigma$ from 0.1 mm to 0.5 both degrades the average field accuracy by around a factor of 4 and increases the spreads of possible field error by a factor of almost 5. 

\section{Conclusion}

In summary, we have demonstrated a robust methodology for assessing the fidelity of different manufacturing processes of complex, non-planar stellarator coil frames for the EPOS device. 
By employing a high-precision laser scanner, we generated a 3D point cloud of the as-manufactured test components. 
This dataset enabled the extraction of two splines along the coil's winding path, which served as the basis for determining the local manufacturing accuracy and quantifying deviations from the intended design.

A comparative analysis revealed performance differences between the two considered design and manufacturing techniques. The wide coil frame produced via CNC machining exhibited superior geometric accuracy, with average deviations being two to three times smaller than those observed on the minimal coil-winding frames produced using AM. 
This underscores the critical role that knowing the manufacturing process plays in achieving the tolerances required for sufficiently accurate stellarator construction.
Having multiple of freedom with  $\sigma$, $L$  and the type of kernel allows to control the smoothness and amplitude of the perturbation functions. This proves the effectiveness of GPs to model manufacturing deviations on non-planar coils. This then provides a strong basis for high fidelity modeling of the magnetic field degradation.
Simulations confirm that maintaining high manufacturing accuracy is crucial, as even small-scale deviations can have a tangible impact on the magnetic field quality. 
At a maximum perturbation amplitude of 0.1 mm, CNC-milling stands as a strong candidate for the manufacturing of the coil frames, yielding high enough field precision to confine electrons and positrons with external coils.
Looking ahead to a reactor-scale stellarator, the challenge of maintaining precision is increased as CNC-milling such a machine is not commercially viable.
Beyond the initial manufacturing, cumulative errors will arise from assembly and alignment, thermal contraction during cryogenic cool-down, and the significant JxB electromagnetic forces that will deform the coils during plasma operations. 
\bibliographystyle{IEEEtran} 
\bibliography{thebibliography}   
\end{document}